\documentclass[twocolumn]{aastex631}

\shorttitle{}
\shortauthors{Campbell et al.}
\graphicspath{{./}{}}

\usepackage{graphicx}
\begin{document}

\title{Application of Deep Learning to the Classification of Stokes Profiles: From the Quiet Sun to Sunspots}

\author[0000-0001-5699-2991]{Ryan J. Campbell}
\affiliation{Astrophysics Research Centre, School of Mathematics and Physics, Queen's University, Belfast, BT7 1NN, UK}

\author[0000-0002-7725-6296]{M. Mathioudakis}
\affiliation{Astrophysics Research Centre, School of Mathematics and Physics, Queen's University, Belfast, BT7 1NN, UK}

\author[0000-0001-5518-8782]{Carlos Quintero Noda}
\affiliation{Instituto de Astrof\'isica de Canarias,
V\'ia L\'actea s/n, E-38205 La Laguna,
Tenerife, Spain}
\affiliation{Departamento de Astrof\'isica, Universidad de La Laguna, E-38206 La Laguna, Tenerife, Spain}

\author[0000-0001-8556-470X]{P. H. Keys}
\affiliation{Astrophysics Research Centre, School of Mathematics and Physics, Queen's University, Belfast, BT7 1NN, UK}

\author[0000-0001-8829-1938]{D. Orozco Suárez}
\affiliation{Instituto de Astrofísica de Andalucía (CSIC), Apdo. de Correos 3004, 18080 Granada, Spain}
\affiliation{Spanish Space Solar Physics Consortium (S3PC), Spain}

\begin{abstract}
The morphology of circular polarisation profiles from solar spectropolarimetric observations encode information about the magnetic field strength,  inclination, and line-of-sight velocity gradients. Previous studies used manual methods or unsupervised machine learning (ML) to classify the shapes of circular polarisation profiles. We trained a multi-layer perceptron (MLP) comparing classifications with unsupervised ML. The method was tested on quiet Sun datasets from DKIST, Hinode, and GREGOR, as well as simulations of granulation and a sunspot. We achieve validation metrics typically close to or above $90\%$. We also present the first statistical analysis of quiet Sun DKIST/ViSP data using inversions and our supervised classifier. We demonstrate that classifications with unsupervised ML alone can introduce systemic errors that could compromise statistical comparisons. DKIST and Hinode classifications in the quiet Sun are similar, despite our modelling indicating spatial resolution differences should alter the shapes of circular polarization signals. Asymmetrical (symmetrical) profiles are less (more) common in GREGOR than DKIST or Hinode data, consistent with narrower response functions in the $1564.85$ nm line. Single-lobed profiles are extremely rare in GREGOR data. In the sunspot simulation, the $630.25$ nm line produces ``double' profiles in the penumbra, likely a manifestation of magneto-optical effects in horizontal fields; these are rarer in the $1564.85$ nm line. We find the $1564.85$ nm line detects more reverse polarity magnetic fields in the penumbra in contradiction to observations. We detect mixed-polarity profiles in nearly one fifth of the penumbra. Supervised ML robustly classifies solar spectropolarimetric data, enabling detailed statistical analyses of magnetic fields.
\end{abstract}

\keywords{Concept: Classification -- Concept: neural networks -- Sun: photosphere --- Sun: magnetic fields}

\section{Introduction} \label{sec:intro}
Solar spectropolarimetry is the technique of measuring and analyzing the polarization of light across different wavelengths in spectral lines that form in different layers of the solar atmosphere. When combined with inversion codes, it is possible to infer physical parameters describing the state of the solar atmosphere \citep{inversions2016}. In addition to applying an inversion code to an entire spectropolarimetric dataset, one can try to understand the nature of the Stokes profiles one is dealing with, including assessing the shapes of the profiles, which can help to determine the appropriate maximum number of free parameters in each atmospheric variable. At first, this was accomplished with a painfully manual processes. For instance,  at a spatial resolution of $0.8-1 \arcsec$, \cite{sig2001} determined the classification and frequency of circular polarisation profiles inside and outside sunspots by determining the signs and amplitudes of Stokes $V$ signals either side of the rest wavelength of the spectral line. This works for most profiles, which either have one negative and positive lobe or have only a single-lobe, with the former further distinguished by whether they are symmetric or asymmetric, but \cite{sig2001} then had to manually validate each profile (among thousands of profiles) to identify the small number that do not fit into either category. The problem with this approach is that as the spatial resolution of the observations increases, the number of profiles that need classified is on the order of millions, not thousands.

Unsupervised machine learning (ML) has become a popular approach for analyzing spectropolarimetric data, especially in the quiet Sun. For example, \cite{khomenko2003} pioneered the application of k-means clustering to study quiet Sun observations in the near-infrared ($1-1.4\arcsec$ resolution), and \cite{Viticchi2011} used similar techniques with Hinode observations at visible wavelengths ($0.32\arcsec$ resolution). While this approach can reveal common shapes of Stokes profiles and their spatial distribution, using k-means clustering to make comparisons between datasets requires caution. This is because the k-means centroid is an “average” profile of each cluster and may not closely resemble any single profile, especially in clusters with high variability. The need to predefine the number clusters, despite clear redundancy and fewer distinct profile categories, underlines the limitations of this method.  Attempts have recently been made to improve upon these issues using shape-based clustering techniques such as k-shape, which refines k-means by focusing on pattern similarity rather than Euclidean distances, making it better suited for grouping Stokes profiles \citep{kshape}.

Unsupervised ML has also been applied to sunspots \citep{Louis2024}. The penumbral magnetic field consists of spines with strong, vertical fields and intraspines with weaker, horizontal fields, forming an interlaced azimuthal structure \citep{borrero2011sunspot}. Sunspots always have a dominant polarity. The dominant polarity magnetic fields (DPMFs) are found throughout, but in the highly complex penumbral regions there can be reverse polarity magnetic fields (RPMFs, sometimes called returning magnetic flux). Simulations by \cite{rempel2012} successfully reproduce observed fine structures, identifying RPMFs in $10\%$ of the penumbra, independent of numerical resolution.  High resolution ($0.2''$) magnetograms by \cite{Langhans2005} initially failed to detect RPMFs, but \cite{franz2013} demonstrated their presence by identifying three-lobed Stokes V profiles using a highly manual method of detection, with $87\%$ of RPMFs associated with downflows, underpinning the importance of spectral resolution and sampling. \cite{cobo2013} further showed by spatially deconvolving Hinode data that detecting RPMFs relies on achieving high spatial resolution. Using degraded synthetic observations produced from simulations, \cite{bharti2019} also concluded that a significant fraction of RPMFs and downflows are hidden in Hinode observations due to a lack of spatial resolution or due to noise. \cite{franz2016} analysed data from both Hinode and GREGOR, finding visible lines to be much more effective at detecting 3-lobed Stokes $V$ profiles, with up to an order of magnitude more, emphasising the importance of the choice of spectral line. Most importantly, high-resolution observations by \cite{scharmer2013}, with a spatial resolution close to 0.15’’, revealed that RPMFs are found only in the deepest layers, with the field weakening at higher layers and expanding over the intraspines. The spatial resolution  enabled a key milestone with their observations showing a RPMF far inside the outer boundary of the penumbra, in key agreement with the simulations of \cite{rempel2012}.

\subsection{Aims and scope}
This study aims to bridge solar physics and machine learning by applying a supervised classification approach to circular polarization profiles from both observed and simulated data. Unlike previous work, which relied on manual or unsupervised methods such as clustering, our approach enables a reproducible, interpretable, and statistically robust comparison of profile morphologies across datasets.

We demonstrate the first application of supervised classification to the morphological analysis of solar Stokes profiles, and in doing so, we present the first statistical analysis of quiet Sun data obtained by the Daniel K. Inouye Solar Telescope's (DKIST's) Visible SpectroPolarimeter (ViSP) instrument. By explicitly defining training labels and evaluating performance on held-out test sets, we ensure rigorous control over classification reliability.

Our primary motivation is to enable consistent and quantitative comparisons between these new DKIST observations and earlier data from telescopes such as Hinode and GREGOR. In addition, we show how the same classification scheme can be extended to sunspot profiles using synthetic observations produced from simulation snapshots. The broader goal is to provide the solar physics community with a validated and reusable supervised classification tool to support future large-scale analysis of high-resolution spectropolarimetric data. The code we developed to classify the DKIST dataset is available on GitHub\footnote{\url{www.github.com/r-j-campbell/StokesClassifier}}.

In Section~\ref{sect:data} we describe the datasets we use, including real observations (Sect.~\ref{sect:observations}) and simulations (Sect.~\ref{sect:simulations}). In Sect.~\ref{sect:methods} we describe the methods we employ, including how the data is preprocessed (Sect.~\ref{sect:preprocessing}), our unsupervised (Sect.~\ref{sect:kmeans}) and supervised (Sect.~\ref{section:MLP}) machine learning methods, our inversion set-up (Sect.~\ref{sect:inversions}) and how we create synthetic observations  (Sect.~\ref{sect:degradation}). In Section~\ref{sect:results} we present our analysis of the data, before discussing the results in Section~\ref{sect:discussion} and suggesting possible future applications in Section~\ref{sect:outlook}.

\begin{table*}
\centering
\begin{tabular}{l|c|c|c}
\hline
\hline
Facility &  DKIST & Hinode  & GREGOR  \\
\hline
Instrument & ViSP & SP & GRIS-IFU \\
Spectral line [{\AA}]& 6302.5 & 6302.5 & 15648.52 \\
Sampling (x, y) [$\arcsec$] & (0.04, 0.0298) & (0.1476, 0.1585) & (0.135, 0.188) \\
FOV (x, y) [$\arcsec$] & 4.26, 76.14 & 162, 302 & 9.024, 12.15  \\
No. frames/repeat scans & 8 & 1 & 70 \\
Spectral sampling [m{\AA}/pix] & 12.85 & 21.5 & 40 \\
Noise level in Stokes $V$ [$I_c$] & $7.5\times10^{-4}$ & $1.1\times10^{-3}$ & $4\times10^{-4}$ \\
\hline
\end{tabular}
\caption{Summary of quiet Sun observations.}
\label{table:observations}
\end{table*}

\begin{table}
  \centering
\caption{Number of nodes employed in the inversions per atmospheric model. Model $1$ is the magnetic atmospheric model. Model $2$ is the non-magnetic atmospheric model. A value of $-1$ indicates that the value in that model was forced to be the same as the other model.}
  \label{tab}
  \begin{tabular}{l   lc lc}
    \hline
    \hline  
    Variable & Model 1 & Model 2\\
    \hline
    $T$ & -1, -1 & 4, 4 \\
    $v_{\rm LOS}$ & 1, 5 & 1, 3 \\
    $B$ & 1, 5 & 0, 0 \\
    $\gamma$ & 1, 5 & 0, 0 \\
    $\phi$ & 1, 1  & 0, 0  \\
    $v_{\rm{mic}}$ & 1, 1 & 1, 1 \\ 
    $v_{\rm{mac}}$ & -1, -1 & 1, 1 \\ 

    \hline
  \end{tabular}
  \label{table:nodes}
\end{table}

\section{Data collection}\label{sect:data}
\subsection{Observations}\label{sect:observations}

We selected three facilities and instruments that are suitable for our analysis, namely the DKIST \citep{DKIST2020} with the Visible Spectropolarimeter (ViSP; \citet{ViSP}), the Solar Optical Telescope (SOT; \citet{SOT}) onboard the Hinode satellite \citep{hinode-solarb} with the SpectroPolarimeter (SP; \citet{hinodesp}), and the GREGOR telescope \citep{Schmidt2012,kleint2020} with the GREGOR Infrared Spectrograph Integral Field Unit (GRIS-IFU; \citet{grisifu}). Each of these facilities are capable of achieving high spatial resolution ($<0.4\arcsec$) with spectrographs that also provide high spectral resolution while recording the full Stokes vector. 
From each of these facilities we select one publicly available dataset. Further details of the ViSP, SP, and GRIS-IFU observations are provided by \cite{campbell2023b}, \cite{suarez2012}, and \cite{campbell2023}. DKIST/ViSP datasets are similar to Hinode/SP in that the instruments are both slit-spectrographs that record the $6301.5/6302.5$~{\AA} line pair. They differ in that DKIST is a ground-based telescope while Hinode is a satellite, such that Hinode/SP observations do not suffer from atmospheric seeing deteriorations. This means that the effective spatial resolution of Hinode is much more stable from one slit position to the next, while this can be highly variable for DKIST/ViSP. The GREGOR/GRIS-IFU observations have a much smaller FOV, but is able to record the spectral information over a 2D FOV strictly simultaneously. Table~\ref{table:observations} summarises the characteristics of the three datasets.

\subsection{Simulations}\label{sect:simulations}
We use 3D snapshots of the magnetoconvection simulation created by \cite{MANCHA2}, computed by the MANCHA code \citep{MANCHA1}. In this simulation the initial seed magnetic field strength is strictly zero, with the Biermann battery effect and local dynamo amplification process generating magnetic fields with strengths close to observations of the quiet Sun. The simulation was also used by \cite{Carlos2023}. Each snapshot has a FOV of $11.52\times11.52$~Mm$^2$ ($1152\times1152$~pixels$^2$) and a mean magnetic field strength at the surface on the order of $100$~G. The simulation box spans from $1$~Mm below the photosphere to $600$~km above it. The pixel size is $10$~km and $7$~km in the horizontal and vertical directions, respectively. We began by selecting one snapshot from within the saturated dynamo regime of the simulation. Later, we added eight additional snapshots with a cadence of $10$~seconds, in order to assess whether the evolution of the classifications was temporally smooth, as expected.

We additionally use a $2000\times2000$~pixel$^2$ segment of a single snapshot of the sunspot simulation from \cite{rempel2012} produced by the MURaM code \citep{vogler2005}. The snapshot is derived from a full simulation which has a domain size of $49.152\times49.152\times6.144$~Mm$^3$, with an initial magnetic field strength at the bottom of $6.4$~kG to $2.56$ kG at the top. The segment has a horizontal extent of $24\times24$~Mm$^2$ with a horizontal (vertical) resolution of $12$~km ($8$~km), and covers a portion of the sunspot umbra, penumbra, and surrounding network-like region. We only make use of a $2.048$~Mm portion in the vertical extent. The time since the initialisation of the simulation is $6$~hours, of which $2.7$~hours were performed in high resolution, and the last $26$~minutes were performed with non-gray radiative transfer, making the snapshot we use suitable for line synthesis.

Both the MANCHA and MURaM snapshots are $3$D domains with the geometric height scale (i.e. in km). We follow the same process as described in \cite{Campbell2021b} to transform  the atmosphere to an optical depth
scale, which is a requirement of the Stokes Inversion based on Response functions (SIR) code \citep{SIR}. This involves using the routine \textit{modelador.x} included in the SIR code.

\subsection{Forward modelling and spatial degradation}\label{sect:degradation}
We made use of the ability to synthesise different spectral lines with SIR. By synthesising Stokes vectors from the MANCHA simulation outputs we are able to investigate the impact of changing the spectral line in isolation on the classification statistics. We synthesised both the $630.25$~nm and $1564.85$~nm lines for both the MANCHA and MURaM simulation snapshots. We always synthesised the spectral lines of interest with a spectral sampling of $12.85$
~m{\AA}. We do not consider any spectral degradation in this study nor do we consider the impact of stray light. The  ability to spatially degrade the synthesised profiles by convolution with a spatial PSF also allows us to investigate the impact of spatial resolution in isolation on the classification statistics for the MANCHA simulation. Beginning with the original numerical resolution of $10$~km in both \textit{x} and \textit{y} directions, we determined the target resolutions of $0.16\arcsec$ and $0.32\arcsec$. The former case is an optimistic average spatial resolution for the DKIST/ViSP dataset, and the latter is the known spatial resolution of Hinode/SP. The spatial degradation was performed by convolving the synthetic observations, for every Stokes parameter and every wavelength, with a 2D Lorentzian kernel, followed by resampling to $0.08\arcsec$pixel$^{-1}$ and $0.16\arcsec$pixel$^{-1}$, respectively for the two target resolutions. 

We do not degrade the synthetic observations produced from the MURaM sunspot simulation because even with a highly optimistic spatial resolution the small amount of inverse magnetic flux disappears.

We always synthesised the Stokes vector with an observation angle, $\mu$, of $1$. We acknowledge that the position of a sunspot on the solar disk would have a significant impact on the morphology of the Stokes $V$ profiles found in the penumbra, but we do not investigate this in this study.

\section{Methods}\label{sect:methods}

\subsection{Pre-processing and data cleaning}\label{sect:preprocessing}
We pre-processed the quiet Sun observational datasets to prepare for unsupervised classification in a number of ways. First, to each observed solar dataset, we applied principal component analysis (PCA) to remove noise. We do not apply this step to synthetic quiet Sun observations as they are noiseless. Second, we attempted a compensation to the relative Doppler shift between Stokes $V$ profiles by determining the shift of Stokes $I$ from the rest wavelength of each profile (i.e., by measuring the minimum intensity across the line profile). Third, we attempted to remove the sign of Stokes $V$ by determining whether the signed maximum value in the blue half of the profile is greater than the signed maximum value in the red half; if the maximum in the blue portion was less than the red the entire profile was multiplied by $-1$. Finally, we normalised every profile by its maximum value. These last three steps are applied to both the observed and synthetic quiet Sun datasets. The result is a population of Stokes $V$ profiles that is better classified according to their shapes rather than their amplitudes, polarities, or Doppler shifts. This pre-processing scheme is similar to that employed by \cite{khomenko2003}.

We do not preprocess the sunspot synthetic observations, except for the normalisation step. The sunspot investigation we present in this study critically relies on determining the location of minority polarity profiles, thus removing the polarity would be inappropriate. Strictly speaking attempting to remove the Doppler shifts even in the case of the quiet Sun is likely unnecessary for supervised machine learning, but in the quiet Sun investigation we wanted to follow the recipe laid out by \cite{Viticchi2011} as closely as possible so that our analysis was comparable. If we did not remove the polarity and Doppler shift, we could not use unsupervised machine learning to select profiles for the training sets for supervised machine learning (discussed in detail in Sect.~\ref{sect:kmeans}).

\subsection{Unsupervised machine learning}\label{sect:kmeans}
We used scikit-learn's \citep{sklearn} implementation of kmeans++ to cluster the circular polarisation profiles in each dataset. Kmeans and kmeans++ are both iterative algorithms used to partition given datasets into a user-defined number of distinct clusters. The algorithms differ only in their initialization procedures.  The kmeans++ algorithm starts by choosing one centroid randomly from the dataset, with subsequent centroids selected based on their distance from the already chosen centroids. The probability of selecting a data point as the next centroid is proportional to its squared distance from the nearest centroid already chosen. This process continues until $K$ centroids are selected, ensuring that they are well spread out across the full dataset. By choosing centroids that are far apart from each other, kmeans++ initialization helps to avoid poor local optima. The centroid selection process was repeated $100$ times for every dataset. The set of initial centroids that results in the lowest inertia (i.e., the clustering with the smallest sum of squared distances), is chosen as the best set of initial centroids. This functionality is included in scikit-learn's default implementation. 

In other applications optimisation of $K$ is important to prevent redundancy between clusters, but since k-means++ is only used in this study for creation of training sets for the supervised machine learning algorithm we only had to make sure $K$ was large enough to provide a representative sample of profiles. The number of clusters, $K$, was set to $35$ for all datasets.  

\begin{figure*}
    \centering
    \includegraphics[width=.9\textwidth]{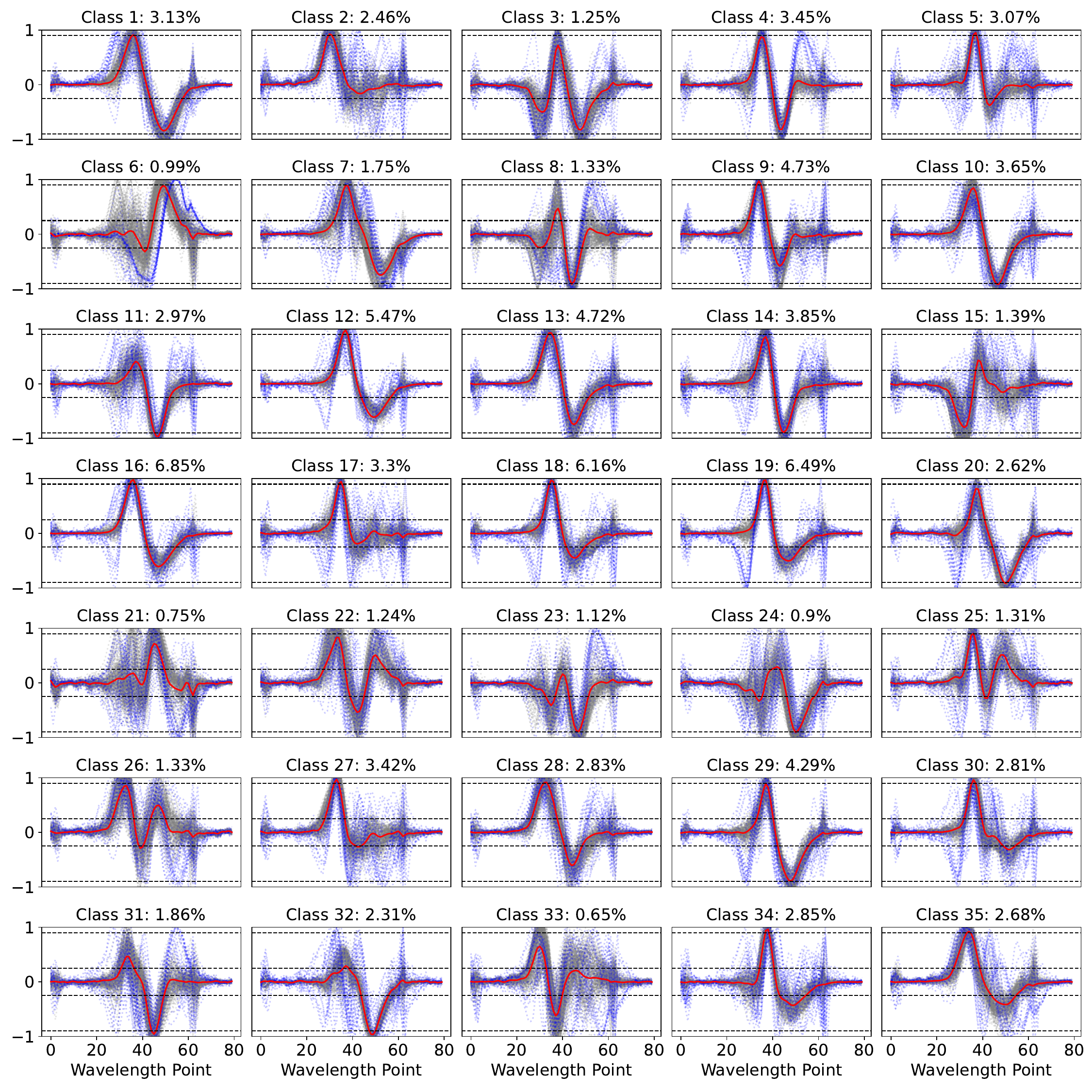}
    \caption{Classes of Stokes $V$ profiles from DKIST/ViSP data for the $6302.5$~{\AA} line from kmeans++. In each panel the centroid is shown in red (\textit{solid lines}), the closest $250$ profiles are shown in gray (\textit{dashed lines}), and the furthest $50$ profiles are shown in blue (\textit{dotted lines}). The percentage of profiles belonging to each class is shown above each panel.}
    \label{fig:kmeans}
\end{figure*}

\begin{figure*}
    \centering
    \includegraphics[width=\textwidth]{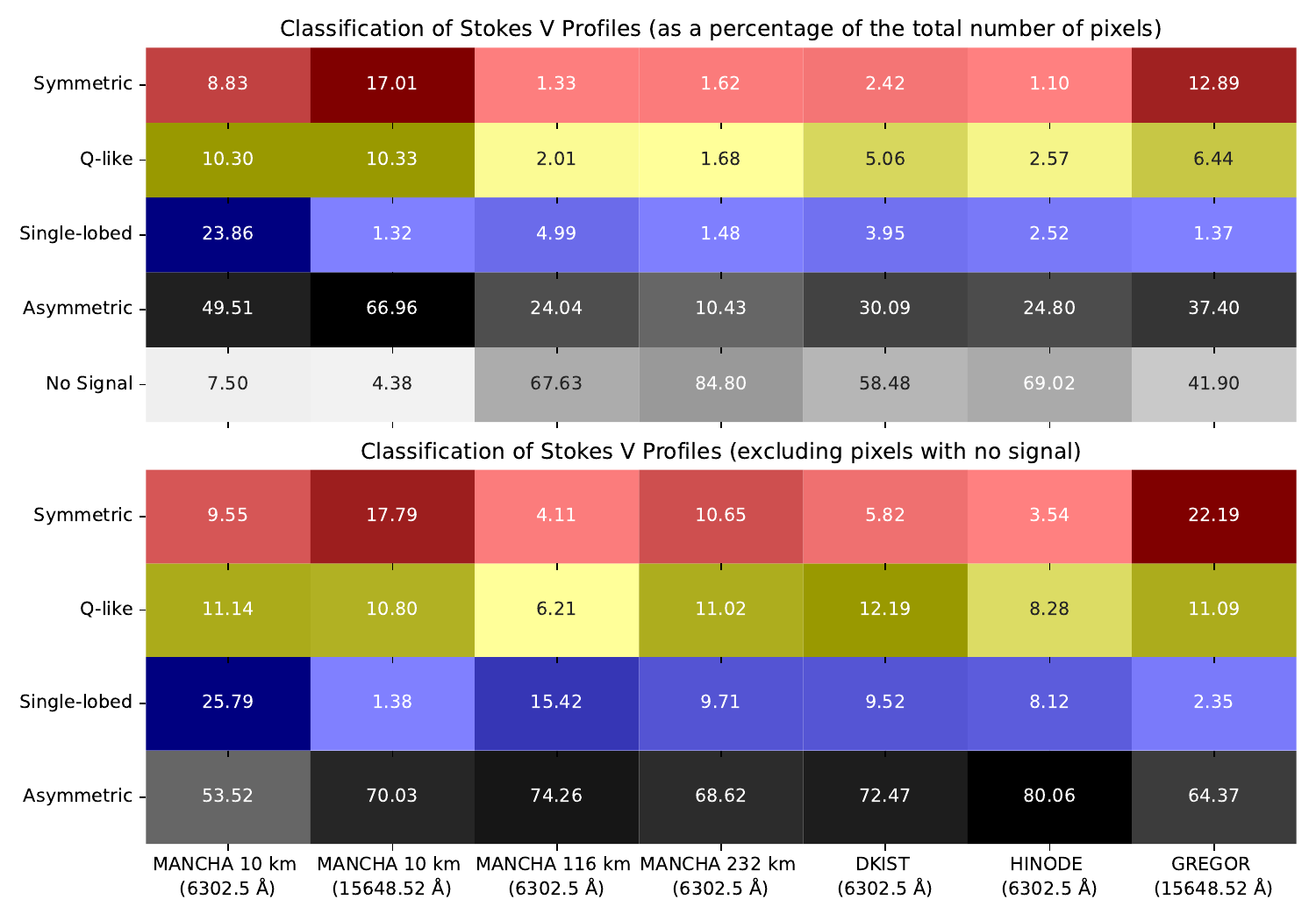}
    \caption{Comparison of classification statistics for circular polarisation profiles from a collection of synthetic (MANCHA simulations) and real (DKIST, Hinode, and GREGOR) quiet Sun observations. Shown are the percentages of Stokes $V$ signals belonging to the four classes (Symmetric, Q-like, Single-lobed, and Asymmetric) as a percentage of the total number of pixels (upper panel) and excluding pixels with no signal (lower panel). For DKIST, Hinode, and GREGOR datasets, the number of pixels with no signal is calculated in Section \ref{section:pol}. }
    \label{fig:classification_stats}
\end{figure*}

\subsection{Supervised machine learning}\label{section:MLP}
We define a multi-layer perceptron (MLP) in the PyTorch framework \citep{pytorch2019} with three fully connected layers, with adjustable input and hidden sizes but a fixed output size. We tested various activation functions (AFs)\footnote{An activation function introduces non-linearity into the neural network, allowing it to learn complex relationships between inputs and outputs.}  but found no clear performance leaders in terms of final accuracies so the AF was always selected as the Swish function\footnote{The sigmoid function maps real-valued inputs to the range (0, 1) and is commonly used in neural networks to introduce smooth non-linearity.}. This has been shown to outperform common AFs like the Rectified Linear Unit (ReLU)\footnote{The Rectified Linear Unit (ReLU) is a common activation function defined as $f(x) = \max(0, x)$, which zeroes out all negative inputs. We do not use this function.} activation function in terms of accuracy and convergence rates \citep{swish}. More importantly, unlike ReLU, Swish smoothly saturates negative values rather than setting negative inputs to zero. Swish is defined as,
\begin{equation}
    f(x) = x \cdot \text{sigmoid}(\beta x),
\end{equation}
where $x$ is the input to the AF. For consistency and simplicity, we kept the $\beta$ hyperparameter in the Swish activation function fixed at 1, ensuring uniform behavior across layers and networks. Since Stokes profiles can be both positive and negative this functionality is essential. We include weight initialization using Xavier uniform initialization\footnote{Xavier uniform initialization sets layer weights to values drawn from a uniform distribution that preserves the variance of activations between layers, helping to avoid vanishing or exploding gradients.} to alleviate the risk of gradient instability during training. We employed the widely used multi-class cross entropy loss function\footnote{Cross entropy loss is a standard loss function for classification tasks that measures the difference between the predicted probability distribution and the true class labels.} with modified class weights to address class imbalance. In simple terms, the class weights are set to be inversely proportional to the class frequencies, such that less frequent classes are assigned higher weights. This enabled us to achieve models with balanced classification performance across all classes. Finally, we chose the widely used Adam optimizer\footnote{The Adam optimizer is an adaptive gradient descent algorithm that adjusts the learning rate for each parameter individually based on estimates of first and second moments of the gradients.} \citep{adam},  incorporated dropout regularization to prevent overfitting, and implemented early stopping to halt training once the validation loss ceased improving.

For each quiet Sun dataset, to create labelled training and validation sets we selected a representative group of profiles using the kmeans++ results and manually labelled thousands of Stokes $V$ profiles. For instance, for the DKIST/ViSP data we created a pool of $8750$ Stokes $V$ profiles by selecting the $250$ closest profiles to each of the $35$ cluster centroids. The rationale is that by drawing the profiles for the training set randomly from this pool we are making sure we train the MLP on the most common types of profiles it can expect to encounter in each dataset and thus hope to achieve a model which generalises well. The size of the pool of profiles was expanded in line with the size of the dataset to be classified to ensure the pool remained truly representative when kmeans++ was applied with $35$ classes. For the sunspot datasets, we found that k-means++ was not producing representative classes because we did not (and could not) remove the polarity. Therefore, when creating training sets we randomly selected and labelled thousands of profiles and before assessing whether enough negative polarity profiles had been included, manually adding a few hundred extra in each case if they were under-represented in the training and validation sets. 

We performed a stratified $70/15/15$ or $80/10/10$ split to divide the labelled profiles into training, validation, and test sets. The training set is used to optimize the model parameters, the validation set is used to tune hyperparameters and monitor for overfitting during training, and the test set is reserved for evaluating the final model's performance on unseen data. Validation and testing scores, along with optimal hyperparameters, can be found in Appendix~\ref{sect:appendix}. 

We parallelized the search across a vast hyperparameter space, which encompassed various parameters such as the sizes of hidden layers (with default options of 128, 256, and 512 neurons in the first layer and 32, 64, and 128 neurons in the second layer), learning rates (ranging from 0.01 to 1 in increments), batch sizes (including 32, 64, 128, 256, and 512), dropout rates (from 0 to 0.5 in increments), and a range of epochs (from 30 to 250). If this search produced unsatisfactory validation metrics, we expanded the parameters and increased the size of the training sets until a satisfactory performance was achieved. Unlike accuracy, which focuses on the overall correct predictions, the f1-score evaluates the trade-off between precision (the proportion of true positives among all positive predictions) and recall (the proportion of true positives among all actual positives). In more simple terms, precision tells us how many of the predicted positive results are correct while recall tells us how many of the actual positive results the model finds. Thus the f1-score balances between getting things right (precision) and finding all the right things (recall). By combining precision and recall into a single metric, the f1-score provides a more nuanced evaluation, particularly useful in scenarios with imbalanced class distributions where both false positives and false negatives are important considerations. Finally. for the model with the highest f1-score, we manually calculated and inspected the class-wise accuracies to make sure the model performed satisfactorily for all classes.

Once the model has been trained and validated, we deploy it to the full dataset. The final output consists of one of the possible predicted labels for each Stokes $V$ profile in the dataset.

\subsection{Spectropolarimetric inversions}\label{sect:inversions} 
We employed the SIR inversion code with a bespoke parallelized wrapper to infer the physical parameters of the atmosphere in every pixel in each quiet Sun dataset. In doing so, we are aiming to determine if there are statistical differences in the atmospheric parameters between the datasets, and also to determine whether there is evidence of gradients in these parameters as a function of optical depth. Every pixel was inverted 15 times each, with randomised initial values in magnetic field strength, $B$, line-of-sight velocity, $v_{\mathrm{LOS}}$, magnetic inclination, $\gamma$ and magnetic azimuth, $\phi$, with the minimum $\chi^2$ solution selected in each case. We made use of abundance data from \cite{asplund}. We made use of the functionality in SIR that allows for automatic selection of nodes up to a maximum, user-defined value. Two cycles per inversion were used, where in the first cycle the temperature, $T$, was the only parameter not constant in optical depth, and in the second additional free parameters were added to $B$, $\gamma$, and $v_{\rm LOS}$. Table \ref{table:nodes} summarises this information. 

\section{Results}\label{sect:results}

\subsection{Fractional polarisation analysis}\label{section:pol}
In order to quantify the fraction of the FOV with confidently measured polarisation signals in each quiet Sun dataset (and thus the fraction with no signal), we counted all those pixels with maximum signal greater than four times the standard deviation (i.e. $4\sigma_n$), as measured in the continuum of Stokes $Q$ and $U$ for linear polarisation, and Stokes $V$ for circular polarisation, across the spectral line. For the DKIST/ViSP data, the fractions of the FOV which have circular and linear polarisation signals greater than the $4\sigma_n$ level are $40.7\%$ and $7.2\%$, respectively. For Hinode/SP, the same quantities are lower at $30.6\%$ and $3.82\%$, respectively. For GREGOR/GRIS, the same quantities are much higher at $60\%$ and $23\%$, respectively \citep{campbell2023}. It is difficult to achieve the signal-to-noise necessary to measure a significantly large fraction of linear polarisation at disk centre in the quiet Sun. As spatial resolution and polarimetric sensitivity improves, it may be possible to classify Stokes $Q$ and/or $U$ profiles, but we decided to focus on Stokes $V$ in this study. Henceforth we analyse only the surviving Stokes $V$ profiles and discard those without sufficient signal. 

\subsection{Unsupervised clustering classification}
Figure \ref{fig:kmeans} shows the classes of circular polarisation profiles returned by kmeans++ when applied to the DKIST/ViSP data. There are four distinct shapes that are immediately evident from the $35$ centroids: asymmetric profiles, with two well-defined lobes (one positive and one negative) but with one dominant lobe (e.g., classes 1, 4, and 5), anti-symmetric or symmetric profiles, with two well-defined lobes (one positive and one negative) with very similar maximum amplitudes (e.g., classes 10, 14, and 29), single-lobed or extremely asymmetric profiles, with only one well-defined lobe (e.g., classes 2, 17, and 21), and Q-like profiles, with up to three lobes, either two positive and one negative, or two negative and one positive (e.g., classes 3, 8, and 22). For Q-like profiles the central lobe may be very close to zero.

Figure \ref{fig:kmeans} also shows the closest $250$ profiles and furthest $50$ profiles in each class. The closest profiles appear very similar in shape to the centroid, but the furthest often bear little resemblance, with additional (or missing) lobes, especially for the most complex classes. 

\begin{figure}
    \centering
    \includegraphics[width=\columnwidth]{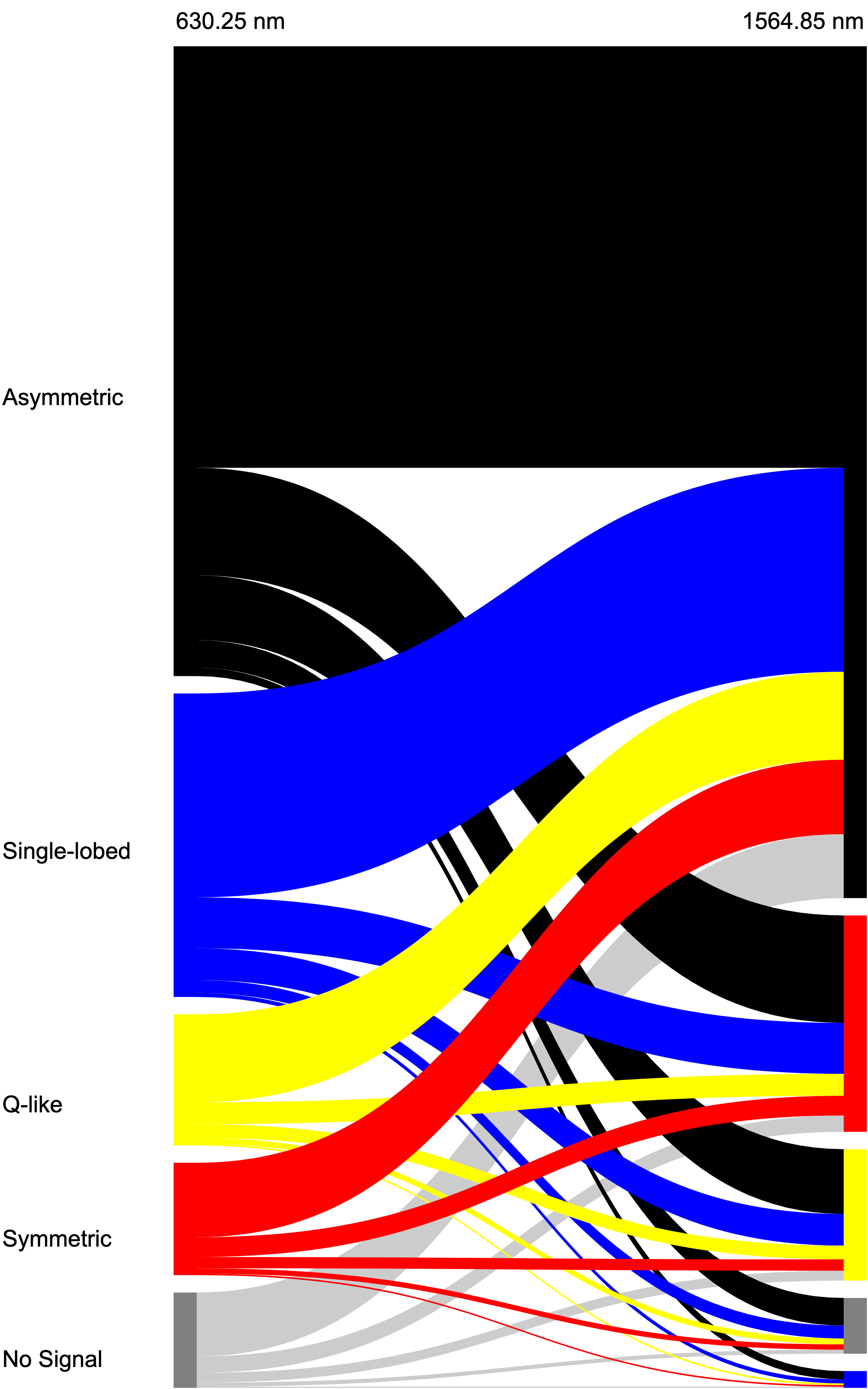}
    \caption{Sankey diagram showing the statistical transfer in the classification statistics produced from the MLP when the original MANCHA quiet Sun data is synthesised in Fe I $630.25$~nm (left) and $1564.85$~nm (right). Numbers for each population are available in the upper panel of Fig.~\ref{fig:classification_stats}. Sankey diagrams are flow-based visualizations that represent how data points move between classes.}
    \label{fig:sankey1}
\end{figure}

\begin{figure*}
    \includegraphics[width=\textwidth]{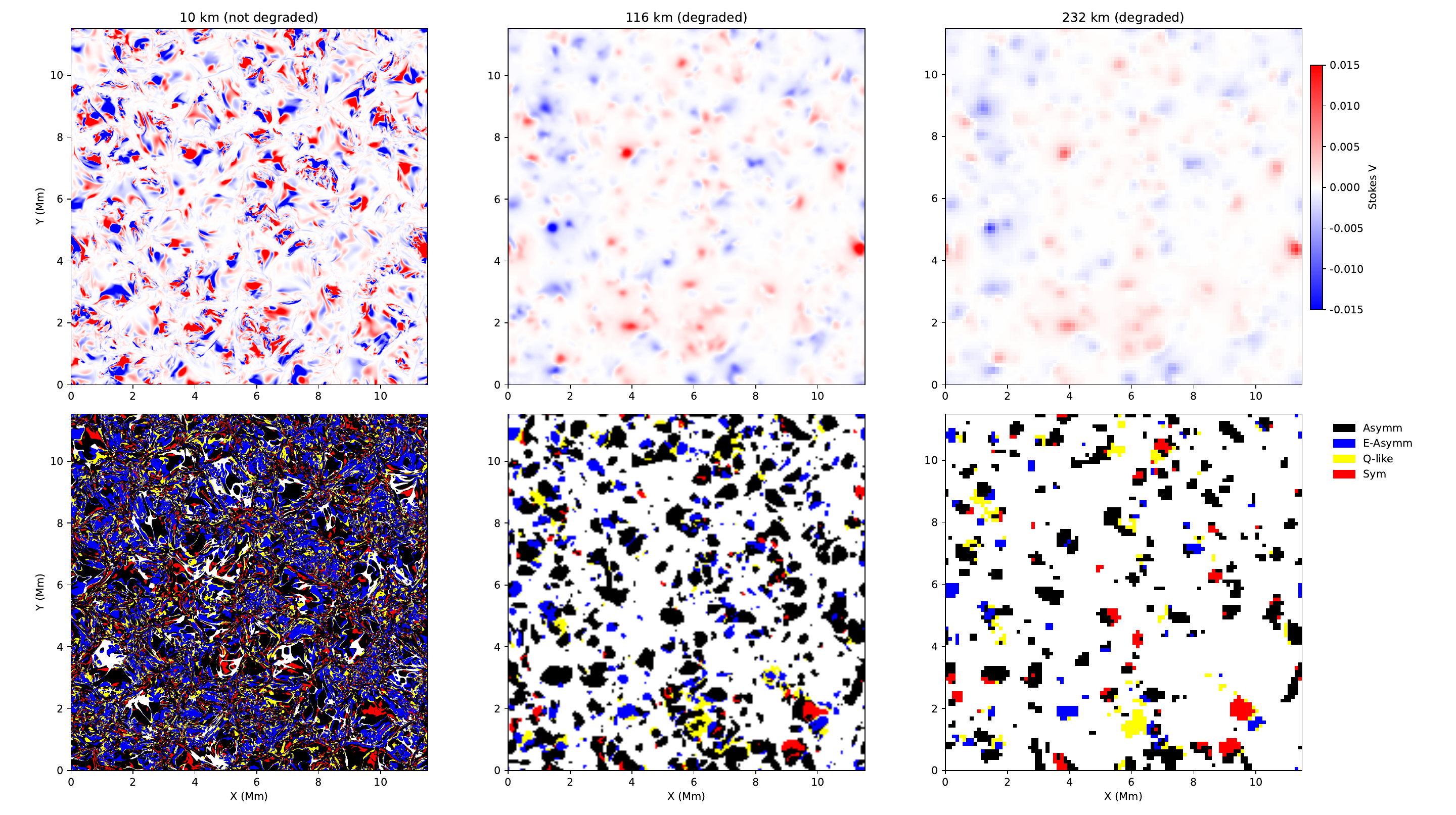}
    \caption{Synthetic Stokes $V$ at a wavelength of $630.24$~nm produced from a MANCHA simulation snapshot (upper row) and the associated labels (lower row) as determined by the MLP. Stokes $V$ profiles are classified as one (E-Asymm), two (Asymm or Sym), or three or more (Q-like) lobes. The results are shown for the 10 km resolution case (without any spatial degradation) and for two lower spatial resolution cases (116 km and 232 km). Stokes $V$ profiles are continuum normalised when synthesised with SIR. An animation showing the temporal evolution over eight timesteps with a cadence of $10$~seconds is available in the online journal.}
    \label{fig:degraded_MANCHA}
\end{figure*}

\subsection{Multi-layer perceptron deployment}\label{section:MLP_results}
The unsupervised machine learning analysis of the quiet Sun data sets identified four distinct categories of Stokes $V$ profiles, which we used to define the labels and the structure of the output layer of the MLP. Specifically, we designed the MLP to take a Stokes $V$ profile as the input and classify it as anti-symmetric, asymmetric, Q-like, or single-lobed. We sometimes use the terms anti-symmetric and symmetric interchangeably. Thus, the output layer of the MLP consists of four nodes, each corresponding to one of the identified profile categories. We note that we took heavy inspiration from the analysis of \cite{Viticchi2011} in determining the labels. The authors used k-means clustering to group profiles into distinct categories based on the morphology of their centroid profiles: network, blue-lobe, red-lobe, Q-like, asymmetric, antisymmetric, and fake. However, we consolidated their red-lobe and blue-lobe into our “single-lobed” or “extremely asymmetric” category, and we did not make a distinction between network and non-network (their network profiles only cover $10\%$ of the field of view, and both we and they normalised the profiles, which  removes the need for this distinction). Finally, when we applied k-means++ to the Hinode, DKIST and GREGOR dataset, with the same number of clusters as \cite{Viticchi2011}, we did not get a fake centroid, so we did not include this category. We also did not encounter these profiles when creating training sets.

To create training and validation sets, we manually labeled thousands of profiles from each dataset (see section \ref{section:MLP} for more details). In Fig. \ref{fig:kmeans}, horizontal lines at $0.25$ and $0.9$ amplitude indicate the thresholds used for classification. For a two-lobed profile to be labelled as anti-symmetric, the subordinate lobe had to reach a minimum amplitude of $0.9$. For a single-lobed classification, a subordinate lobe could not exceed an amplitude of $0.25$. Otherwise, the two-lobed profiles in each of these cases would be considered asymmetric. Finally, Q-like profiles are distinct enough to be labelled with ease. Although these threshold values are arbitrary, they were introduced to maintain consistency in the labelling process, which remains subject to human error.

Figure \ref{fig:classification_stats} shows the classification statistics for a number of quiet Sun datasets from GREGOR, Hinode, and DKIST (see Section \ref{sect:observations}) and from MANCHA simulations (see Section \ref{sect:simulations}), both at the original numerical resolution ($10$~km) and degraded to $116$~km and $232$~km resolutions (see Section \ref{sect:degradation} for a description of the degradation process).

Comparing the classification statistics for the MANCHA simulation, synthesised in the visible and near infrared (NIR) line and not degraded, it is clear that the choice of spectral line has a significant impact. The vast majority of pixels have detectable signal in both cases ($92.5\%$ for the visible lines versus $95.62\%$ for the NIR). Comparing the statistics as a percentage of the total number of pixels versus excluding pixels with no signal therefore makes little difference. Considering the former, symmetric ($8.83\%$ versus $17.01\%$) and asymmetric ($49.51\%$ versus $66.96\%$) profiles are significantly more common when observed in the NIR line, while single-lobed profiles are significantly less common ($23.86\%$ versus $1.32\%$). The number of Q-like profiles is similar ($10.30\%$ versus $10.33\%$).

Since there are significant differences in the classification statistics when the original MANCHA cube is synthesised in Fe I $630.25$~nm and $1564.85$~nm, the question naturally follows where the transfer is occurring between labels (and, thus, which profiles are changing shapes). Figure~\ref{fig:sankey1} illustrates the statistical transfer in classification results from the MLP model, comparing outputs when the original MANCHA cube is synthesized in Fe I $630.25$~nm (left) and $1564.85$~nm (right), with widths between the two populations representing the proportion of each classification's transfer between categories. The value in representing the data in this way is immediately evident because it is clear there is significant transfer between labels in each case. For instance, one might expect the $17.45\%$ difference in the number of asymmetric profiles for the $1564.85$~nm line relative to the $630.25$~nm line to be explainable by this exact number, or close to it, being drawn from other categories. However, from Fig.~\ref{fig:sankey1} one can observe that this increase is driven by a much larger inflow occurring towards the asymmetric ($1564.85$~nm) category, as well as a significant outflow from this category. The largest donor to the asymmetric ($1564.85$~nm) category is from single-lobed ($630.25$~nm) profiles, while the largest loss from the asymmetric ($630.25$~nm) category is towards symmetric ($1564.85$~nm) profiles. Notably, every category experiences some degree of exchange, underscoring the complexity of the changes in the shapes of the Stokes $V$ profiles. Perhaps another unexpected observation is the not insignificant number of Q-like ($630.25$~nm) profiles that become asymmetric ($1564.85$~nm) and symmetric ($1564.85$~nm) profiles. The collapse in the number of single-lobed ($630.25$~nm) profiles from the second most populus label to the lowest (even below no signal in $1564.85$~nm) occurs in the manner one would expect - the largest transfer is towards asymmetric ($1564.85$~nm) profiles as the secondary lobe increases in amplitude in the more magnetically sensitive near infrared line, with a still significant but lower transfer towards symmetric ($1564.85$~nm) profiles.

The first observation to be made when comparing the statistics for the MANCHA simulation with the Fe I $630.25$~nm spectral line in the cases before and after degradation to $116$~km and $232$~km spatial resolutions, is the massive collapse in the number of pixels that have detectable signals. Here, we have defined detectable signals as at least three times the $1\sigma$ noise level of the DKIST/ViSP observations (i.e. $2.25\times10^{-3}$$I_c$). Of course, it is possible to measure weaker signals, but we defined it this way as it enables comparison with the DKIST data. Before degradation, $92.5\%$ of pixels have detectable signals, but degrading to $116$~km and $232$~km lowers this to $32.37\%$ and $15.2\%$, respectively. Figure~\ref{fig:degraded_MANCHA} shows maps of the Stokes $V$ profiles for both the degraded and original cases, along with the classification labels for each pixel. The reduction in amplitude and in the fraction of the FOV that has detectable signals is striking. Of the surviving profiles, fewer of the degraded profiles are single-lobed, and more are asymmetric, when one considers the classification statistics as a percentage excluding those pixels with no signal. Notably, there is a higher fraction of Q-like profiles in the original case than the $116$~km degraded case, but for the $232$~km degraded case the number of Q-like profiles is similar to the original case (around $11\%$). One might expect Q-like profiles to be generated by the mixing of Stokes $V$ signals that occurs due to the convolution with the spatial PSF, and this can be observed to occur in specific regions of the simulation. One might also expect single-lobed profiles to be generated by the collapse in amplitude, but we do not find evidence for this.

\begin{figure*}
    \centering
    \includegraphics[width=\textwidth]{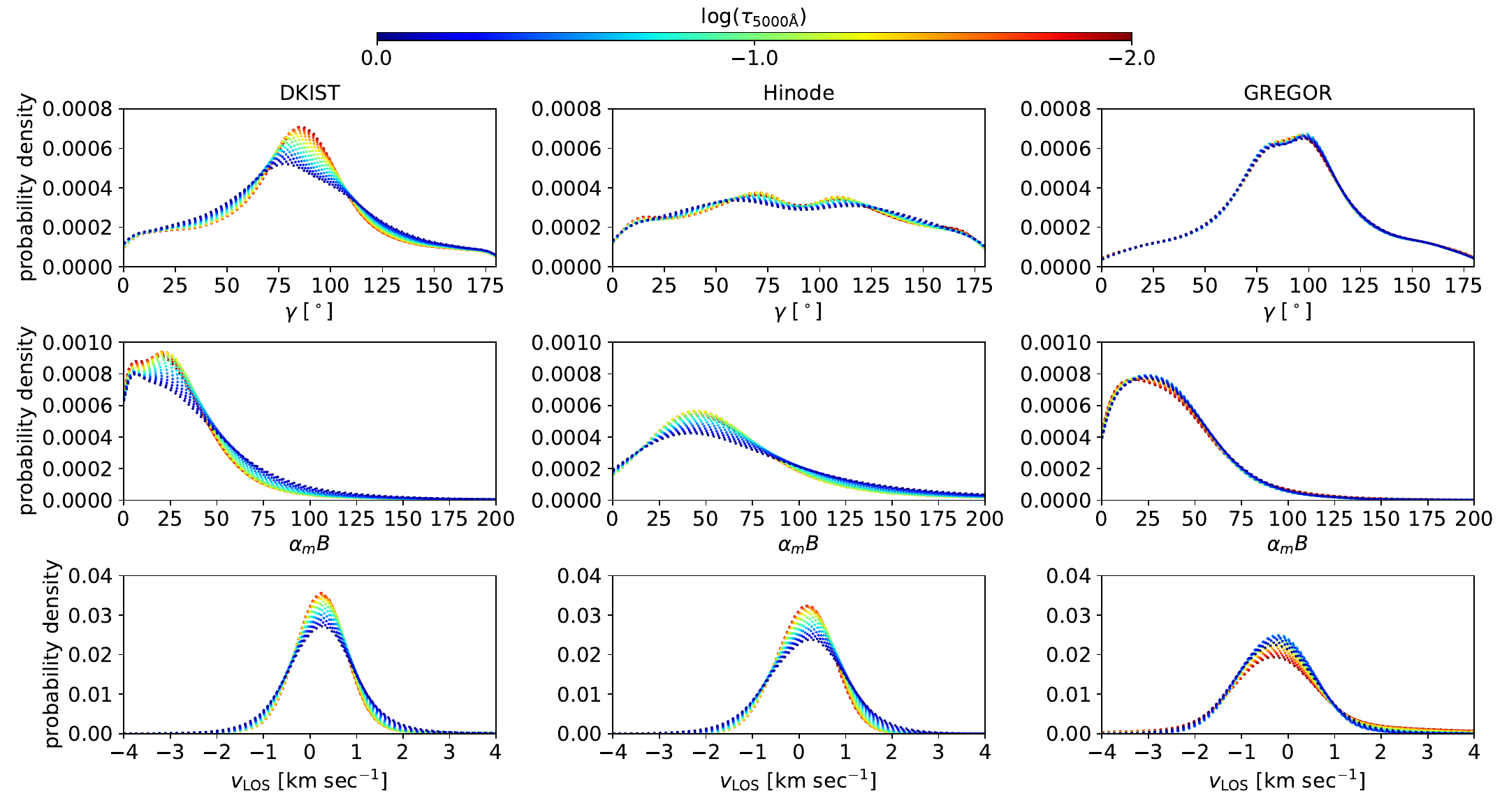}
    \caption{Probability density distributions of $\gamma$ (top row), $\alpha_m B$ (second row), and $v_{\mathrm{LOS}}$ (third row) returned from the second cycle of the inversions (as described in Table~\ref{table:nodes}). The distributions are shown for the three quiet Sun datasets (as described Table~\ref{table:observations}). The distributions are shown for increments in the logarithm of the optical depth at $5000$~{\AA}, log($\tau_{5000\text{\AA}}$), of $0.1$ between $0.0$ and $-2.0$. Pixels that had no measured polarization in either Stokes $Q$, $U$, or $V$ are excluded.}
    \label{fig:inversions}
\end{figure*}

\begin{figure*}
    \centering
    \includegraphics[width=\textwidth]{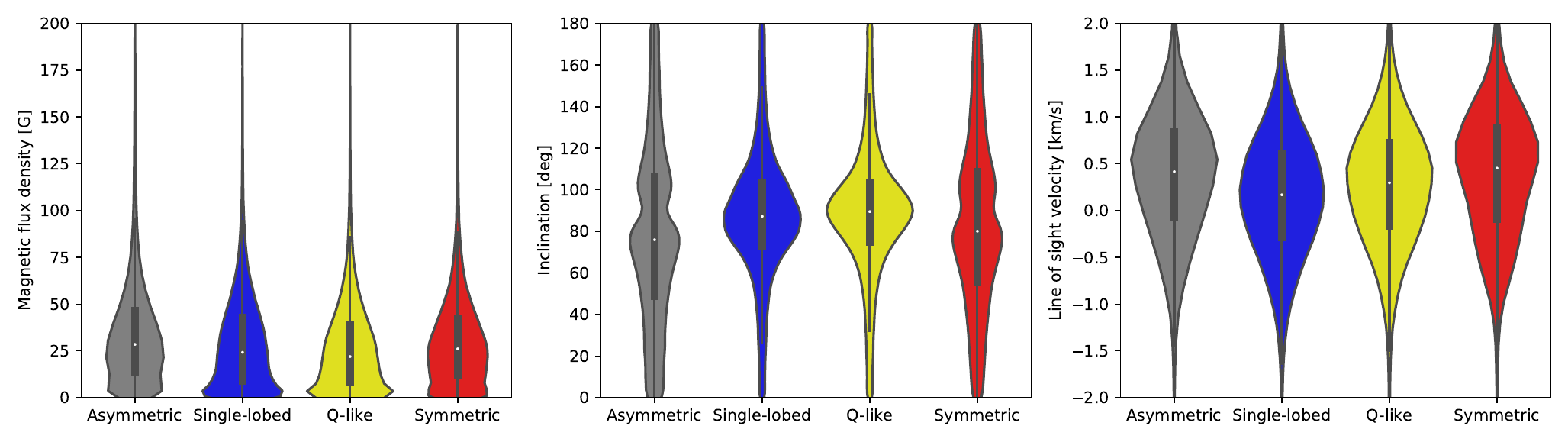}
    \caption{Violin plots showing the distribution $\alpha_m B$, $\gamma$, and $v_{\mathrm{LOS}}$ from DKIST/ViSP inversions at log($\tau_{5000\text{\AA}}$)~$= - 1.0$, grouped by Stokes $V$ morphological class as determined by the MLP. Each violin shows the kernel density estimate for pixels in a given class, with the white dot marking the median and the thick bar the interquartile range.}
    \label{fig:inversions_MLP}
\end{figure*}

\begin{figure}
    \centering
\includegraphics[width=1.0\linewidth]{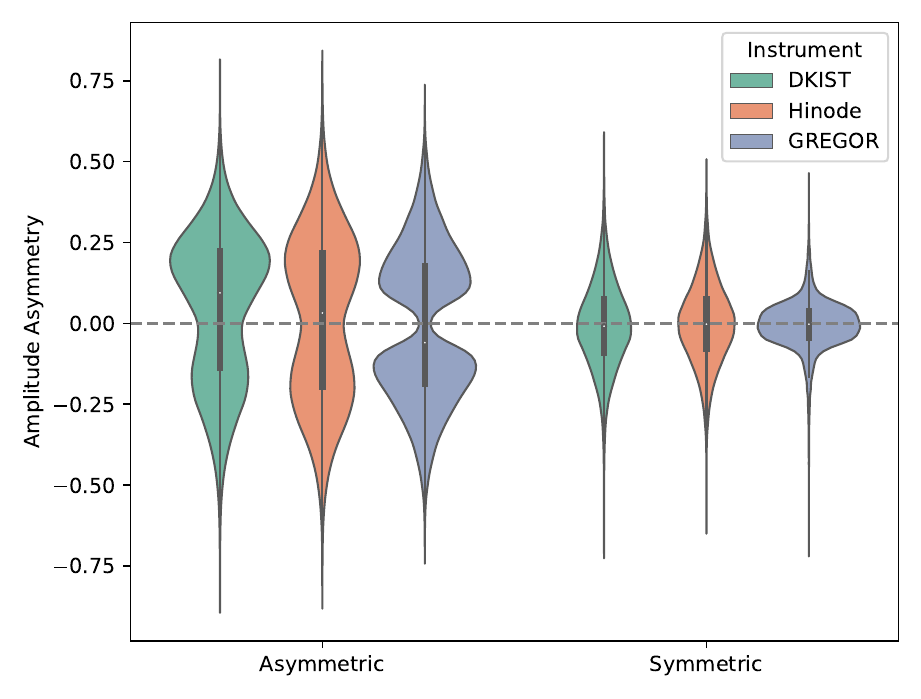}
    \caption{Violin plots of amplitude asymmetry ($\delta a$) for circular polarisation profiles classified as asymmetric and symmetric, shown separately for the DKIST, Hinode, and GREGOR datasets. Each violin represents the distribution of $\delta a$ within a given class and instrument. Horizontal dashed lines indicate zero amplitude asymmetry. Distributions are plotted with inner box plots to indicate medians and interquartile ranges.}
    \label{fig:asym_violin_plot}
\end{figure}

Comparing the classification statistics for the two degraded resolutions ($116$~km and $232$~km) could theoretically be insightful because if there are significant differences between the higher and lower resolution cases this could help when comparing observations from different telescopes with different effective spatial resolutions, or even datasets from the same ground-based facilities and instruments but with different seeing quality. There is only a small difference in the percentage of Stokes $V$ profiles with signal having asymmetric shapes ($74.26\%$ versus $68.62\%$), but the percentage with single-lobed profiles was observed to decrease ($15.42\%$ versus $9.71\%$) while the number of Q-like ($6.21\%$ versus $11.02\%$) and symmetric ($4.11\%$ versus $10.65\%$) profiles both increased. Importantly, this analysis does indicate that the spatial resolution should have an impact on the shapes of observed circular polarisation profiles. However, if one examines the statistics for Hinode and DKIST, which both observed with the same Fe I $630.25$~nm spectral line and would be expected to have significant differences in effective spatial resolution, this does not seem to be the case. The biggest statistical differences between the two datasets is in the percentage of asymmetric profiles, which is greater for Hinode by $7.59\%$. However, from our modelling with the degraded synthetic observations produced from the MANCHA simulation it would have been predicted that the opposite would be true.

The classification statistics for GREGOR are significantly different than for DKIST and Hinode, with fewer asymmetric and single-lobed profiles and many more symmetrical profiles. This result is perhaps explainable due this line having a high excitation potential and its formation within a narrow range of optical depths. This leads to less asymmetric Stokes $V$ profiles due to a smaller range of optical depths in which there can be gradients in the atmospheric parameters influencing the circular polarisation profiles \citep{dan2002,franz2016}. The classification statistics for the GREGOR datasets differ from the DKIST and Hinode statistics in the same way that the original MANCHA simulation differs when synthesised in the Fe I $630.25$~nm and Fe I $1564.85$~nm lines (i.e. more symmetrical profiles and fewer single lobed profiles). 

\subsection{Statistical analysis of inversions}

The DKIST and GREGOR inclination distributions shown in Fig.~\ref{fig:inversions} are similar, with Hinode being significantly less likely to have an inclination around $90^\circ$. This can be understood in the context of signal to noise, spatial resolution, and their impacts on the fraction of polarisation measured (see section \ref{section:pol}). Ultimately, Hinode/SP is inferior to both GREGOR and DKIST in terms of the fraction of linear polarisation measured, resulting in a less horizontal inclination distribution. However, the GREGOR inclination distribution does vary in one significant way - the GREGOR distribution shows no variance with optical depth, while the DKIST distribution shows significant variance. This analysis also applies to the magnetic flux density distributions, given that the DKIST and GREGOR distributions peak at similar (lower) values, while the Hinode distribution is on average stronger. Since GREGOR and DKIST observed with a lower noise level, these observations accessed weaker magnetic fields than the Hinode scan. However, the magnetic flux density is also a function of the magnetic filling factor, $\alpha_m$, and thus the spatial resolution, which should be expected to be significantly different between DKIST and Hinode given DKIST's much larger aperture. The line-of-sight velocity distributions for DKIST and Hinode are very similar, while the GREGOR distribution varies less with optical depth. 

We also ran the inversions with a node scheme where $T$ was not forced to be the same in both model atmospheres. This significantly increases the number of free parameters, but nevertheless we found that it made no difference to the $B$ distribution, which showed almost no additional variance.

To investigate whether the inferred atmospheric parameters differ systematically between Stokes $V$ morphological classes, we examined the distributions of $\alpha_m B$, $\gamma$, and $v_\mathrm{LOS}$ for the DKIST dataset (see Fig.~\ref{fig:inversions_MLP}). We find no statistically significant differences in magnetic flux density across classes; the distributions are broadly similar. Inclination angles, however, show a clear trend: asymmetric and antisymmetric profiles are more likely to be associated with vertical fields, whereas Q-like and extremely asymmetric (single-lobed) profiles are more commonly linked to nearly horizontal fields ($\gamma \approx 90^\circ$). In terms of $v_\mathrm{LOS}$, asymmetric and antisymmetric profiles are slightly redshifted on average, which is consistent with those profiles being located in intergranular lanes, while Q-like and single-lobed profiles are more narrowly distributed around zero velocity.

\subsection{Comparing supervised and unsupervised classification}
\begin{figure}
    \centering
    \includegraphics[width=\columnwidth]{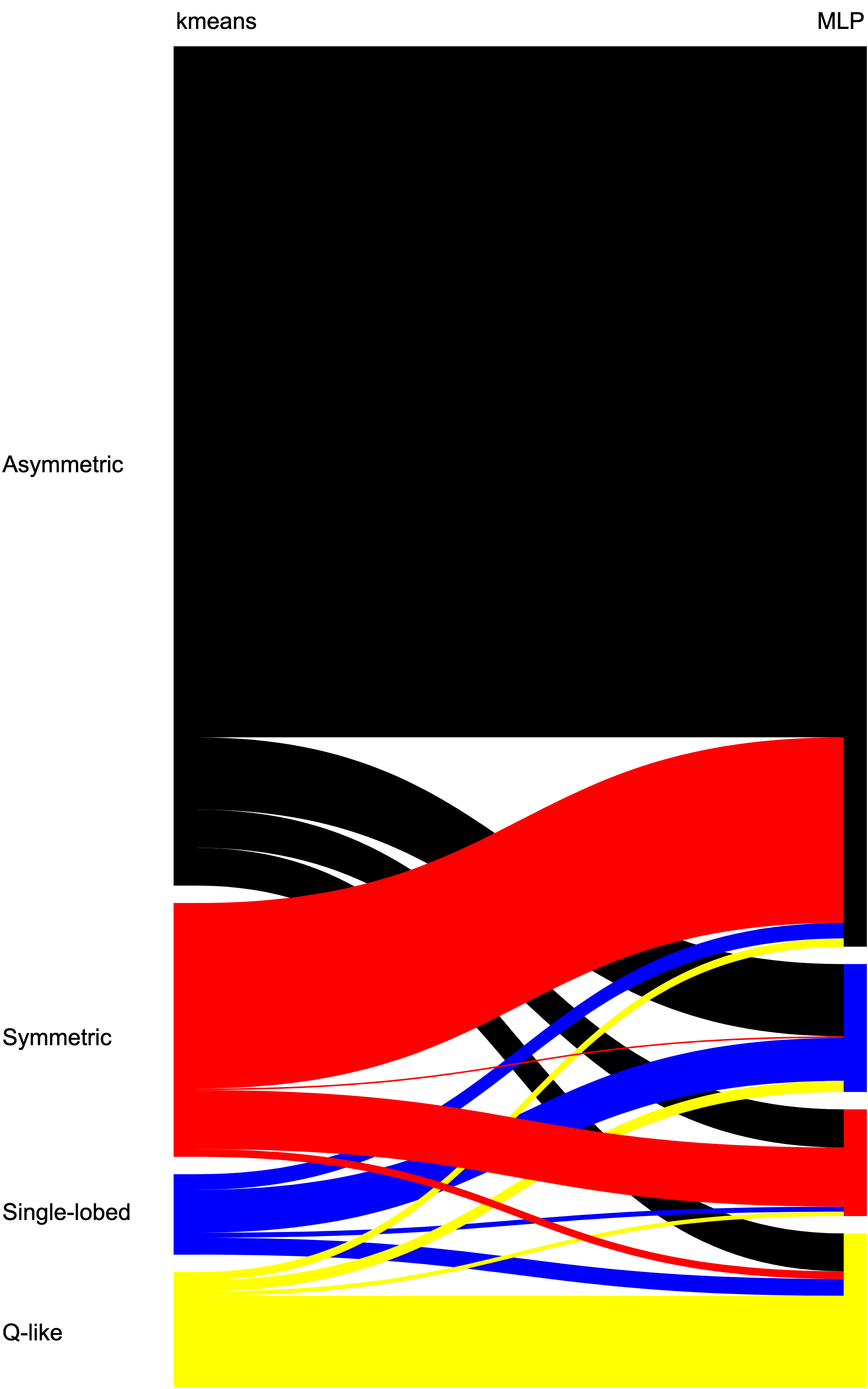}
    \caption{Sankey diagram showing the statistical difference in the classification statistics produced from k-means ($left$) and the MLP ($right$) for Stokes $V$ profiles from DKIST/ViSP quiet Sun data.}
    \label{fig:sankey2}
\end{figure}

From Fig.~\ref{fig:kmeans} it is clear that the assumption, when considering k-means$++$ classification, that every profile in a given class can be categorized according to the shape of the centroid profile, is probably false. But the question remains how severe the errors are that are introduced. In order to determine this, we used the same upper ($0.9$) and lower ($0.25$) normalised amplitude limits we used to label Stokes $V$ profiles for creating the training and validation sets, but instead labelled the centroid profiles for all $35$ classes for the DKIST dataset. For instance, for Class 3 in Fig.~\ref{fig:kmeans}, the centroid profile is clearly a Q-like profile, so all profiles assigned to that cluster were labelled as Q-like. Figure~\ref{fig:sankey2} illustrates the results of this classification analysis compared to the classification statistics produced by the MLP. Evidently, there are differences between the two methods.  From the perspective of taking the MLP results as the ground truth, k-means$++$ clustering seems to have significantly over-estimated the number of symmetric profiles. There is also a less significant misclassification of single-lobed profiles as asymmetric profiles. The process was repeated assuming $70$ clusters instead but this did not impact the results in any significant way, indicating that $35$ clusters is already redundant. In other words, this systematic error is not removed by increasing the number of clusters, and it is not an optimisation problem - instead, it is a problem inherent to the k-means$++$ clustering approach. Since most of the centroid profiles would be classified as asymmetric profiles, it logically follows that most misclassifications would involve profiles being assigned to asymmetrical centroids.

The existence of systematic errors with the pure k-means$++$ classification method is problematic because unless the classification of all datasets are biased in the same way, a robust classification that allows for comparing datasets (as we have done in Figure.~\ref{fig:classification_stats}) is impossible. With the MLP approach, we are not only able to benchmark the process with a target validation and test score, but also to implement class-wise validation, so any errors impact the categories to similar extents. Ultimately, this means when comparing classification statistics for different datasets, as long as the overall validation scores, test scores, and class-wise validation scores are similar, as is the case in all analysis we present in this paper, we can argue we have produced a more robust classification overall.

\subsection{Statistical analysis of amplitude asymmetries}

To further characterise the morphology of Stokes~$V$ profiles classified by our supervised method, we examined their amplitude asymmetries.  Following standard definitions \citep[e.g.,][]{solanki1984}, the amplitude asymmetry is given by \begin{equation} \delta a = \frac{|a_b| - |a_r|}{|a_b| + |a_r|}, \end{equation} where $a_b$ and $a_r$ are the extrema of the blue and red lobes, respectively.

In Figure~\ref{fig:asym_violin_plot}, we show violin plots of $\delta a$ for the asymmetric and symmetric classes in the DKIST, Hinode, and GREGOR datasets. As expected, symmetric profiles exhibit near-zero amplitude asymmetries, while the asymmetric profiles show a  distribution that peaks at non-zero (positive and negative) values, providing empirical evidence that the MLP classifier has successfully distinguished between these morphological types based on asymmetry. The DKIST distribution of asymmetric profiles is skewed toward one sign of $\delta a$, whereas the Hinode distribution appears more balanced; this may be due to the smaller field of view of the DKIST scan. The narrower interquartile range in the GREGOR distributions may reflect the narrower response function of the Fe I $1564.85$~nm line, which is less responsive to the atmospheric gradients in velocity or magnetic field that produce large amplitude asymmetries.

\subsection{Detection of reverse polarity in sunspots}
\begin{figure}
    \centering
    \includegraphics[width=\columnwidth]{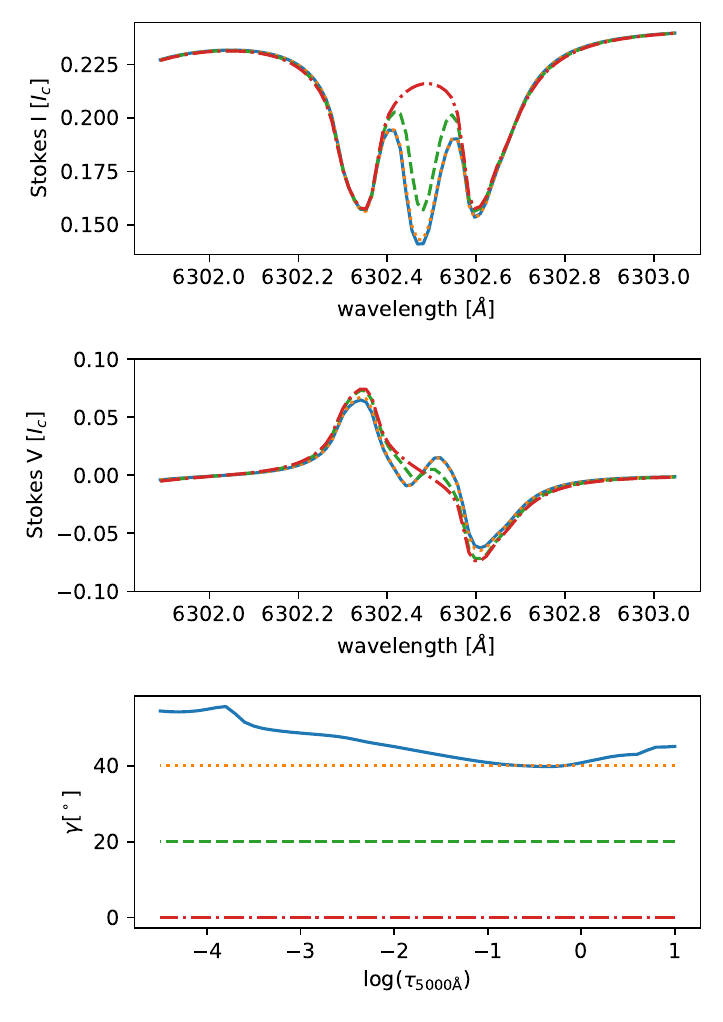}
    \caption{Sample double profile synthesised from the associated original model atmosphere (\textit{solid, blue lines}), shown with the synthesised profiles with the inclination adjusted to be constant in optical depth at values of $40^\circ$ (\textit{orange, dotted lines}), $20^\circ$ (\textit{dashed, green lines}), and $0^\circ$ (\textit{dot-dashed, red lines}). The Stokes $I$ (\textit{upper panel}) and $V$ (\textit{middle panel}) profile are shown for the Fe I $630.25$~nm line along with the inclination (\textit{lower panel}). }
    \label{fig:double_prof}
\end{figure}

\begin{figure*}
    \centering
    \includegraphics[width=\textwidth]{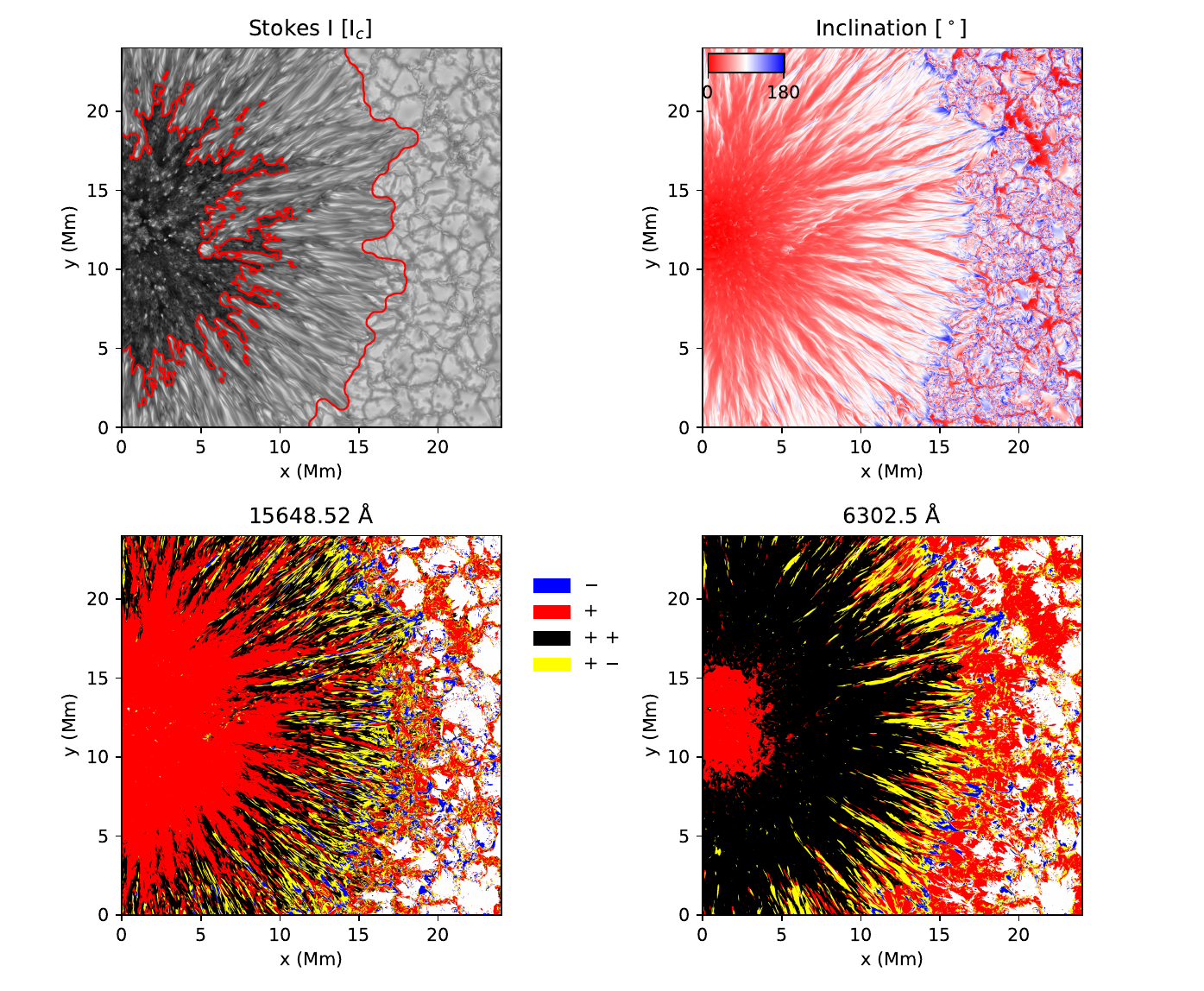}
    \caption{Classification statistics shown for the MURaM sunspot simulation. Shown is the synthetic Stokes $I$ (\textit{upper left}) at a continuum wavelength, the magnetic inclination angle, $\gamma$, at an optical depth of log$(\tau_{5000\text{\AA}}) = -0.5$ (\textit{upper right}), and the classification labels for the near infrared (\textit{lower left}) and visible (\textit{lower right}) lines as determined by the MLP. Stokes $V$ profiles are classified as positive ($+$, \textit{red}), negative ($-$, \textit{blue}), mixed polarity or other ($+-$, \textit{yellow}) and double positive ($++$, \textit{black}). Pixels without maximum signal amplitudes greater than $0.02$ $I_c$ are excluded (shown in \textit{white}). }
    \label{fig:sunspot_MLP_maps}
\end{figure*}

The four labels we used for the quiet Sun do not fully capture the diversity of sunspot circular polarisation profiles. The data pre-processing steps are also not appropriate, except the normalisation step, as we want circular polarisation profiles of different polarities to be distinguished by the MLP. After carefully examining thousands of randomly selected profiles produced from the MURaM sunspot simulation described in Section \ref{sect:simulations}, we determined that there are five common types of profiles. The first two classes are two-lobed, positive or negative polarity profiles. A positive polarity Stokes V profile is a two-lobed profile where the blue lobe is more positive and the red lobe more negative, indicating a magnetic field pointing outward ($\gamma < 90^\circ$ at disk centre), while a negative polarity profile has the red lobe more positive and the blue lobe more negative, indicating a field pointing inward ($\gamma > 90^\circ$ at disk centre).
In both cases the lobes may have asymmetries in area or amplitude but not more complex shapes. The next two classes are similar, except that they have four lobes instead of two. They have the dominant positive and negative lobe, but two other positive or negative lobes closer to the line core wavelength than the two dominant lobes. We refer to these profiles as double profiles. The final class is the mixed-polarity case, where the dominant polarity is unclear, which is typically the result of a highly complex stratification in the inclination angle with optical depth, especially one which crosses $90^\circ$ once or more in the photosphere.

The positive polarity profiles are very common in the dataset for both lines, but selecting negative polarity double profiles is challenging due to their very low occurrence. Despite extensive hyperparameter tuning and testing we were unable to accurately train the MLP to locate negative double profiles, so instead we maintained positive double profiles as a distinct label while encapsulating both simple negative profiles and negative double profiles into a single label.

Before analysing where double positive profiles (henceforth collectively referred to as double profiles) are located, and determining how common they are, we need to establish the atmospheric conditions which create them.  Figure~\ref{fig:double_prof} shows a sample $630.25$~nm double profile from the penumbra. It is worth pointing out that very close inspection of the Stokes $V$ profile in the $1564.85$~nm line also shows two additional lobes near the line centre, but the amplitude is so small relative to the dominant lobes that it would easily neglected. We changed the inclination in the model, first to make it invariant with optical depth, and set $\gamma = 40^\circ$ at all optical depths. This value was chosen as this is approximately equal to the value at log($\tau_{5000\text{\AA}}$~$=-0.5$). We then re-synthesised the Stokes vector and found that the Stokes $V$ and Stokes $I$ profiles did not change. This already rules out inclination gradients as being responsible for this double profile. We then adjusted the inclination again to make the model more vertical, setting it to $\gamma = 20^\circ$ and $\gamma = 0^\circ$ at all optical depths, and re-synthesised the Stokes vector in each case. This is also shown in Fig.~\ref{fig:double_prof}. We find that as the model becomes progressively more vertical, the double profile shape of Stokes $V$ disappears, leaving a much simpler positive profile. We also note that the shape of the Stokes $I$ profile changes substantially at the line core. It would of course be possible to train the MLP to analyse Stokes $I$, and with an understanding of how the shapes of the profiles are formed, this indicates it would be possible to locate and differentiate between highly inclined and more vertical locations purely from analysing the shape of the intensity profiles. From the analysis of Stokes $V$, which is the focus of this study, one can deduce that we might expect double profiles to be located in the most magnetically inclined locations (i.e. in the penumbra). We also systematically adjusted the magnetic field strength, line-of-sight velocity, and magnetic azimuth angle in order to make sure there were no degenerate solutions that could be responsible for the double profile in this case. We found that only the inclination was responsible.

As a final test before proceeding with the classification of the $630.25$~nm and $1564.85$~nm lines, we also synthesised the Fe I $525.02$ nm line for the same model used in the sample shown in Fig.~\ref{fig:double_prof}. This line has the same effective Land\'e g-factor as the NIR line, but according to the response functions is more sensitive to changes in inclination than both the $630.25$~nm and $1564.85$~nm lines \citep{quintero2021}. As expected, the relative amplitude of the minor lobes in the Stokes $V$ profile was largest in the $525.02$ nm line.

Figure~\ref{fig:sunspot_MLP_maps} shows maps of the classification results for both the Fe I $630.25$~nm and $1564.85$~nm lines. It is clear that the central umbral region is distinct from the much more varied and complex penumbral region. In the $630.25$~nm line, in the sunspot, positive profiles dominant in the central umbral region, but double positive profiles become dominant in the outer umbral region and are also common in the penumbra along the more vertical filamentary structures. In the $1564.85$~nm line, positive profiles dominate almost the entire umbra, extending along the more vertical filamentary structures deep into the penumbra. Mixed polarity profiles extend in the other direction in both lines, from the sunspot moat toward the umbra, and they are found in the same regions in the $1564.85$~nm and $630.25$~nm lines. The vast majority of negative profiles are found in the sunspot moat or surrounding network areas. Indeed, the network areas are dominated by positive, mixed polarity, and negative profiles. In the $1564.85$~nm line there are some locations outside the sunspot where double positive profiles can be found but they are rare. The locations of each the positive and negative profiles are consistent with expectations from the inclination angle (i.e. negative profiles are found in those regions with an inclination angle close to $180^\circ$ at log$(\tau_{5000\text{\AA}}) = -0.5$. However, negative profiles can be found in the middle and inner penumbra in the $1564.85$~nm line but not the $630.25$~nm line.

\begin{figure}
    \centering
    \includegraphics[width=\columnwidth]{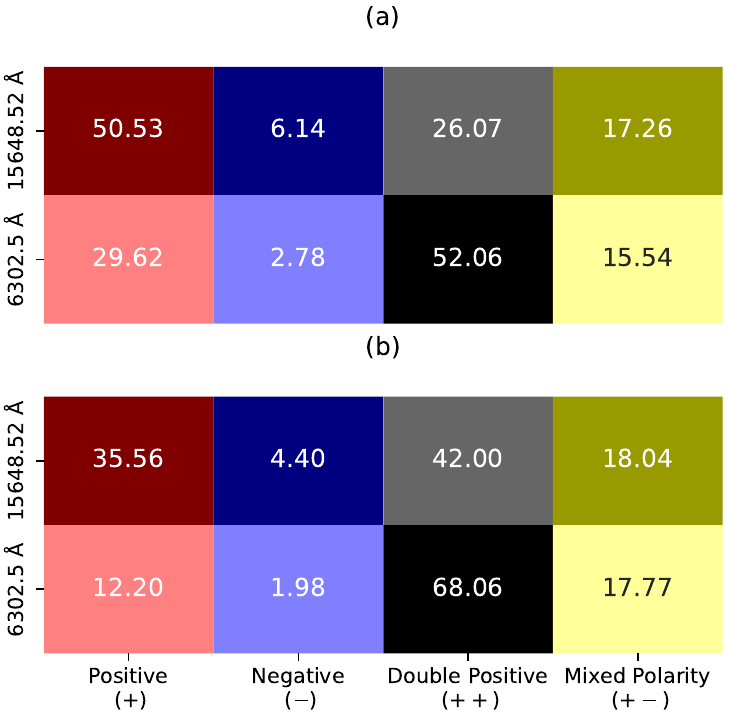}\newline
    \caption{Comparison of classification statistics for circular polarisation profiles synthesised from the MURaM sunspot simulation snapshot in Fe I $630.25$~nm (\textit{lower row}) and $1564.85$~nm (\textit{upper row}). Shown are the percentages of Stokes $V$ signals belonging to the five classes (Positive, Negative, Double Positive, and Mixed Polarity) excluding those with maximum signal amplitudes lower than $0.02$ $I_c$. Plot (a) shows the statistics for the full sunspot segment while (b) shows only for the penumbra isolated as indicated by the boundary in Fig.~\ref{fig:sunspot_MLP_maps}.}
    \label{fig:sunspot_MLP_stats}
\end{figure}

Figure~\ref{fig:sunspot_MLP_stats} shows the tabulated classification results for both the Fe I $630.25$~nm and $1564.85$~nm lines in the full FOV. It is clear that the number of profiles with DPMFs for both lines is very similar, as for the $630.25$~nm line the total is $81.68\%$ being classified as positive or double positive, while for the $1564.85$~nm lines it is $76.60\%$. The total number of profiles with RPMFs (i.e. negative profiles) is over twice as large for the $1564.85$~nm line ($6.14\%$) than the $630.25$~nm line ($2.78\%$), but ultimately profiles with a RPMF are rare in both cases. The number of mixed polarity profiles for both lines is similar.  Figure~\ref{fig:sunspot_MLP_stats} also shows the statistics for the penumbra alone. For both lines, there are relatively more double positive profiles, because of the more inclined magnetic fields, which dominate the statistics especially for the $630.25$~nm line, and negative profiles are relatively even rarer for both lines. 

\subsection{Disentangling polarity reversals from magneto-optical effects}
When choosing labels for the sunspot circular polarisation profiles, we implicitly made the assumption that highly complex profiles with multiple lobes and thus an ambiguous polarity (which we organised into the mixed polarity class) were likely the result of a highly complex stratification in $\gamma$ along the line of sight, especially one which crosses the polarity inversion line at $90^\circ$. However, magneto-optical effects are a competing mechanism that can cause the morphology of Stokes $V$ to deviate from being two-lobed \citep{wittmann}. To determine whether so-called polarity reversals along the line of sight are truly responsible for the Stokes $V$ morphologies in the sunspot penumbra, we examined the underlying atmospheric model parameters used to synthesize the Stokes vectors. We computed, at each pixel, whether the inclination angle crossed $90^\circ$ along the log$(\tau_{5000\text{\AA}}) = [0.0, -2.0]$ range, indicating a reversal in the direction of the magnetic field relative to the observer. As a robustness check, we repeated this analysis using the sign of the LOS magnetic field ($B_z$) and found the statistics to be virtually unchanged. We then computed the fraction of pixels in each morphological class that coincide with a polarity reversal, restricting the analysis to the penumbra and to pixels with significant Stokes $V$ signal in the $630.25$~nm and $1564.85$~nm lines.

Across the entire penumbra, we find that polarity reversals occur in  $25\%$ of pixels in the optical depth range log$(\tau_{5000\text{\AA}}) = [0.0, -2.0]$. We find that, for the $630.25$~nm and $1564.85$~nm lines respectively,  $74.7\%$ and $67.2\%$ of mixed-polarity profiles are associated with a polarity reversal along the line of sight. In contrast, only $8.3\%$ and $20.1\%$ of double-positive profiles exhibit a polarity reversal. This indicates that mixed-polarity profiles are most often associated with genuine polarity reversals in the atmosphere, while the double positive profiles are more likely to originate from magneto-optical effects in strongly inclined fields. Although magneto-optical effects and polarity reversals can both contribute to the same Stokes $V$ profile, the clear statistical trends observed across these classes suggest that the MLP classifier is capturing meaningful differences in the underlying magnetic structure. Furthermore, depending on the chosen metric (accuracy or f1-score; validation or test set), classification performance varies between approximately $0.78$ and $0.94$ (see Appendix \ref{sect:appendix}). Nonetheless, the trends observed here demonstrate that the MLP has successfully learned to separate Stokes $V$ morphologies in a way that corresponds to the underlying dominant physical drivers in the atmosphere.

\section{Discussion}\label{sect:discussion}
DKIST and Hinode have very similar classification statistics across all four classes, even though the degraded MANCHA synthetic observations indicate that differences in spatial resolution should generate significant differences in the classification statistics. This is also the case despite there being some measurable differences in the statistical properties of the quiet Sun regions each telescope observed as inferred from the inversions (namely, the DKIST and GREGOR datasets have weaker and more horizontal fields than the Hinode dataset). In order for this experiment to be more robust the two facilities would have to observe the same region of the solar surface at the same time, but the result is surprising nonetheless. The MLP classifier separates profiles into classes with distinct atmospheric properties: asymmetric and antisymmetric profiles are preferentially associated with vertical magnetic fields and slight redshifts (i.e. in intergranular lanes), while Q-like and single-lobed profiles correspond to more horizontal fields and velocities near zero on average.

The statistics for the original MANCHA simulation when synthesised with the two lines differ in the same way as DKIST/Hinode ($630.25$~nm) and GREGOR ($1564.85$~nm). This lends weight to the interpretation that the intrinsic properties of the $1564.85$~nm line are responsible for these differences, however the original MANCHA simulation and the GREGOR datasets have very similar classification statistics, despite having very different spatial resolutions. Overall the lack of variance with optical depth for the near infrared lines, in all three parameters shown in Fig.~\ref{fig:inversions}, can be understood as a manifestation of the more narrow response functions. Importantly, this is consistent with the interpretation of the classification statistics in Section~\ref{section:MLP_results}, where fewer asymmetric (and more symmetric) profiles were found by the MLP in the GREGOR datasets than in the DKIST or Hinode datasets. Additionally, this indicates that inversion schemes where gradients are not permitted in optical depth are more appropriate for the NIR doublet than for the visible doublet. Across all three observations, amplitude asymmetries are near-zero for symmetric profiles and significantly non-zero for asymmetric profiles, providing empirical support that the classifier has separated morphologies in a physically meaningful way. We observe that the DKIST distribution is skewed toward one sign of $\delta a$, potentially due to its limited field of view, while the narrower interquartile range in the GREGOR results may further reflect the reduced sensitivity of the Fe I $1564.85$~nm line to the line-of-sight velocity gradients that produce large asymmetries. Because amplitude asymmetry is only well-defined for two-lobed profiles, using the MLP classifier to isolate symmetric and asymmetric cases before computing $\delta a$ ensures that the calculation is applied only where physically meaningful, avoiding spurious contributions from Q-like or single-lobed profiles.

We demonstrated that using k-means to determine the shapes of the profiles can result in systematic errors if one assumes the centroid profile is truly representative of all profiles in a given cluster. The supervised ML approach still has errors, especially in terms of the validation and test accuracy, which is never $100\%$ (see Appendix~\ref{sect:appendix}), human error in creating the labelled training and validation sets, poorly defined labels, and lack of proper generalisation to unseen data. However, it avoids the redundancy involved in having a much larger number of classes or centroids than necessary, and avoids the risk of introducing systematic errors generated from assuming a centroid is truly representative of every profile in the cluster. As we have also shown for the sunspot, it does not necessarily require as much preprocessing to remove Doppler shifts and polarities as unsupervised ML.

The biggest difference between the $1564.85$~nm and $630.25$~nm spectral lines in the sunspot relates to higher prevalence of double positive profiles in the latter compared to the former. The presence of additional lobes near the line center in Stokes $V$ is ultimately a manifestation of magneto-optical effects, which become increasingly significant as the magnetic field inclination deviates from the longitudinal direction \citep{dan2002, rolf}. This effect is more pronounced in the $630.25$ nm line, whose response function of Stokes $V$ to changes in inclination is larger than that of the $1564.85$~nm line \citep{quintero2021}.

We find evidence that the $630.25$ nm line is less capable than the $1564.85$~nm line at detecting RPMFs in sunspots. The fact that mixed polarity profiles (which are typically interpreted as indicating the presence of RPMFs in real observations) are $2.8-5.6$ times more common than negative profiles underlines the complexity of the penumbral magnetic field structure, and this is without any spatial PSF applied to the synthetic observations. Our results agree with analysis of observations by \cite{franz2016} who found mixed-polarity profiles to be significantly more common than negative polarity profiles in GREGOR and Hinode data. Our results are also consistent with \cite{franz2013}, who found $4\%$ of the penumbral area harboured RPMFs, while up to $17\%$ could harbour RPMFs if 3-lobe profiles (i.e. mixed polarity profiles) are included in the analysis. 

\cite{franz2016} indicated that the reason RPMFs are more commonly observed in Hinode observations than GREGOR observations might be because of intrinsic properties of the spectral lines combined with RPMFs being located in downflows. Flux tubes in downflows can become evacuated, with the NIR line less likely to observe RPMF profiles as it is sensitive to deeper layers. It is true that the response function of the $1564.85$~nm line to $B$ and $\gamma$ generally peaks deeper in the atmosphere, but the difference is small, and response functions are ultimately model-dependent and specific to each Stokes parameter \citep{quintero2021}.  In direct contradiction to observational results, our analysis indicates that the $1564.85$~nm line is more capable at detecting RPMFs in the middle and inner penumbra. Further, in the simulations, there is visibly more RPMF in the penumbral filaments as one moves to deeper optical depths, so it is not clear that this explanation is entirely satisfactory. This indicates that currently the MURaM simulation does not align which the observations in terms of the height (or optical depth) that RPMFs are manifesting. We suggest the difference might be more as a result of the differences in spatial resolution in real observations, and ultimately conclude that very high spatial resolution is essential for understanding the nature of the sunspot penumbra. In this respect, future DKIST observations with good seeing conditions have an important role to play.

\section{Outlook}\label{sect:outlook}
The supervised machine learning method we have devised could be applied to many more use cases. For instance, it can be used to located so-called blue- or red-``humped'' profiles in sunspots that have been identified as locations where magnetic reconnection may have occurred \citep{franz2013, Hamedivafa2024}. It could be used to ascertain whether the quiet Sun DKIST/ViSP dataset analysed in this paper has an effective spatial resolution that is significantly worse than initially estimated or anticipated. When additional ViSP observations are obtained with good seeing conditions, the new dataset could be classified and compared to the existing one, with a significant difference expected if a higher effective spatial resolution is achieved at the same signal-to-noise. The method could be used to analyse plage regions to hunt for the existence of emergent opposite polarities at potential sites of magnetic reconnection \citep{chitta2019}. With these and many other possible use cases in mind, we have made the code for the MLP classifier  freely available on \hyperlink{https://github.com/r-j-campbell/StokesClassifier}{GitHub}.

\facilities{DKIST \citep{DKIST2020, ViSP}, Hinode \citep{hinode-solarb, SOT, hinodesp}, GREGOR \citep{Schmidt2012,kleint2020,grisifu}.}
\software{PyTorch \citep{pytorch2019}, Astropy \citep{astropy3}, Matplotlib \citep{mpl}, Numpy \citep{numpy}.}

 The observational data used during this research is openly available. Readers can access DKIST data from the \hyperlink{https://dkist.data.nso.edu}{DKIST Data Centre Archive} under proposal identifier pid$\_1\_36$, Hinode data from the \hyperlink{http://www2.hao.ucar.edu/csac}{Community Spectropolarimetric Analysis Center}, and GREGOR data from the \hyperlink{https://archive.sdc.leibniz-kis.de/}{KIS Science Data Centre}.

We thank the anonymous referee to helped us improve the manuscript. We would like to thank Jose Carlos del Toro Iniesta, Rolf Schlichenmaier, and Javier Trujillo Bueno for their very insightful discussions of magneto-optical effects in sunspots. RJC and MM acknowledge support from STFC (ST/P000304/1, ST/X000923/1) and the EU Horizon 2020 program (SOLARNET, 824135).  The 1.5-meter GREGOR solar telescope was built by a German consortium under the leadership of the Institute for Solar Physics (KIS) in Freiburg with the Leibniz Institute for Astrophysics Potsdam, the Institute for Astrophysics Göttingen, and the Max Planck Institute for Solar System Research in Göttingen as partners, and with contributions by the Instituto de Astrofísica de Canarias and the Astronomical Institute of the Academy of Sciences of the Czech Republic. GREGOR's redesign was carried out by KIS whose technical staff is gratefully acknowledged.
Hinode is a Japanese mission developed and launched by ISAS/JAXA, collaborating with NAOJ, NASA and UKSA. Support for the post-launch operation is provided by JAXA and NAOJ (Japan), UKSA (U.K.), NASA, ESA, and NSC (Norway). The research is based in part on data collected with the Daniel K. Inouye
Solar Telescope (DKIST), a facility of the National Solar Observatory (NSO). NSO is managed by the Association
of Universities for Research in Astronomy (AURA), Inc., and is funded by the National Science Foundation (NSF). 
Any opinions, findings and conclusions or recommendations expressed in this publication are those of the
authors and do not necessarily reflect the views of the NSF or AURA. DKIST is located on land of spiritual and cultural significance
to Native Hawaiian people. The use of this important site to further scientific knowledge is done so with
appreciation and respect. We use data provided by M. Rempel at the National Center for Atmospheric
Research (NCAR). The NCAR is sponsored 
by the NSF. 

 \appendix
\section{Validation metrics}\label{sect:appendix}
\begin{table}[h]
\centering
\caption{Validation and test metrics (including accuracy and f1-score) and best hyperparameter combinations (including hidden layer (HL) sizes, learning rates (LRs), batch sizes (BS), dropout rates (DRs), and number of epochs.}
\begin{tabular}{lccccccccc}
\toprule
Dataset & HLs & LRs & BS & DRs & Epochs & Val. A. & Val. F1 & Test A. & Test F1 \\
MANCHA 10~km ($630.25$~nm) & 512-32 & 0.03 & 256 & 0.0-0.0 & 105 & 0.90 & 0.89 & 0.88 & 0.88\\
MANCHA 10~km ($1564.85$~nm) & 512-32 & 0.005 & 256 & 0.0-0.0 & 170 & 0.91 & 0.88 & 0.90 & 0.88 \\
MANCHA 116~km ($630.25$~nm) & 128-64 & 0.01 & 128 & 0.0-0.1 & 185 &  0.96 & 0.93 & 0.89 & 0.84 \\
MANCHA 232~km ($630.25$~nm) & 256-64 & 0.01 & 128 & 0.0-0.1 & 112 & 0.94 & 0.90 & 0.91 & 0.85 \\
DKIST ($630.25$~nm) & 128-128 & 0.01 & 125 & 0.0-0.0 & 125 & 0.94 & 0.90 & 0.94 & 0.91 \\
HINODE ($630.25$~nm) & 256-128 & 0.01 & 256 & 0.0-0.0 & 160 & 0.96 & 0.92 & 0.93 & 0.86 \\
GREGOR ($1564.85$~nm) & 128-64 & 0.01 & 128 & 0.0-0.0 & 50 & 0.90 & 0.90 & 0.90 & 0.90\\
MURaM 12~km ($630.25$~nm) & 1024-64 & 0.05 & 512 & 0.3-0.0 & 140 & 0.94 & 0.90 & 0.82 & 0.78 \\
MURaM 12~km ($1564.85$~nm) & 128-256 & 0.005 & 512 & 0.0-0.0 & 175 & 0.87 & 0.86 & 0.84 & 0.83 \\
\end{tabular}
\end{table}

\bibliography{sample631}{}
\bibliographystyle{aasjournal}

\end{document}